\documentclass[showpacs,12pt,a4paper]{article}  % СËÄ12pt           %¡¾1¡¿
\usepackage{geometry}
\usepackage{cite}  %¡¾revtex4-1 Åųâciteºê°ü£¡£¡£¡¡¿¡¾4¡¿
\usepackage{amsmath,amssymb}
\usepackage{latexsym,amscd,amsthm,amsxtra,graphicx,mathrsfs,amsfonts}  %\mathsrc,\mathfrak ¶ÔÓÚÏ£À°×Öĸ²»Æð×÷Óã¡
\usepackage{verbatim}

\usepackage{dcolumn}% Align table columns on decimal point
\usepackage{hyperref}

\usepackage{color}

\usepackage{multirow}
\usepackage{extarrows}
\usepackage{empheq,mdframed,lipsum}
\usepackage{diagbox}

\usepackage{caption,epsfig,subfig}

\usepackage{dsfont}
\usepackage{array}
\usepackage{booktabs}

\usepackage{xpatch}
\usepackage{blindtext}

\usepackage{CJK}
\usepackage{CJKfntef}
\usepackage{bm}
\usepackage{cases}

\usepackage{simplewick}   %%¡¾ÓÃÓÚWick¶¨ÀíËã·ûÊÕËõ¡¿
\usepackage{pdfpages}

\newcommand{\be}{\begin{equation}}
\newcommand{\ee}{\end{equation}}
\newcommand{\ba}{\begin{eqnarray}}
\newcommand{\ea}{\end{eqnarray}}
\newcommand{\bea}{\begin{equation}\left\{ \begin{aligned}}
\newcommand{\eea}{ \end{aligned} \right. \end{equation}}
\newcommand{\bc}{\begin{numcases}  }
\newcommand{\ec}{\end{numcases}   }
\newcommand{\non}{\nonumber}

\newcommand{\ben}{\begin{equation*}}
\newcommand{\een}{\end{equation*}}
\newcommand{\ban}{\begin{eqnarray*} }
\newcommand{\ean}{\end{eqnarray*}}

\allowdisplaybreaks[2]

\begin{document}

%\begin{CJK*}{GBK}{song}

%\preprint{APS/xxx-xxx}

\title{Compact $N$-quark Hadron Mass Dependence on $N^4$:\\
A Classical Field Picture}
%\thanks{xxxxxxxxxxxxxxxxxxxx}

\author{
Rui-Cheng LI\thanks{rui-chengli@163.com} \\
{\small\it School of Physical Sciences, University of Chinese Academy of Sciences, Beijing 100049, China}
}

\begin{comment}
\author{Rui-Cheng LI}
\email{rui-chengli@163.com}
\affiliation{School of Physical Sciences, University of Chinese Academy of Sciences, Beijing 100049, China\\}
\end{comment}

\maketitle

%\date{\today}% It is always \today, today,
             %  but any date may be explicitly specified

\begin{abstract}
We give a hypothesis on the mass spectrum of compact $N$-quark hadron states in a classical field picture, which indicates that there would be a mass dependence on about $N^4$. We call our model ``bag-tube oscillation model", which can be seemed as a kind of combination of quark-bag model and flux-tube model. The large decay widths due to large masses might be the reason why the compact $N$-quark hadrons still disappear so far.
\begin{description}
\item[PACS numbers]\quad    11.15.Kc, \quad 12.39.Ba, \quad 12.39.Mk, \quad  12.39.Pn, \quad  12.40.Yx,  \quad 12.90.+b
\item[Key words]\quad multi-quark state,\quad constituent quark,\quad glueball,\quad bag model,\quad flux-tube model,\quad hadron mass spectrum
\end{description}
\end{abstract}

%11.15.-q   Gauge field theories
%¡¾11.15.Kc¡¿   Classical and semiclassical tech.
%11.90.+t   Other topics in general theory of fields and particles
%¡¾12.39.Ba¡¿   Bag model
%¡¾12.39.Mk¡¿   Glueball and nonstandard multi-quark/gluon states
%¡¾12.39.Pn¡¿   Potential models
%12.40.-y   Other models for strong interaction
%¡¾12.40.Yx¡¿   Hadron mass models and calculations
%¡¾12.90.+b¡¿   Miscellaneous theoritical ideas and models

%\maketitle
%%¡¾LiRC¡¿Note that in REVTeX 4.1 the abstract must be specified before the \maketitle command and there is no need to embed it in an explicit minipage environment.

%\tableofcontents

\thispagestyle{empty}
%\pagestyle{empty}  % for all pages!
%%%%%%%%%%%%%%%%%%%%%%%%%%%%%%%%%%%%%%%%%%%%%%%%%%%%%
\newpage
\setcounter{page}{1}

\section{\label{Introduction}Introduction}

It is reasonable and necessary to go back to
the classical field picture from the quantum field picture.
The classical field picture has been applied into the renormalization in quantum field theory.
For example, the physical mass of an electron in the quantum electrodynamics (QED) was seemed as the combination
of the bare mass of electron and the effective mass of electric field surrounding the electron (or, the self-energy correction of electron).
Similarly for a quark,
besides of the bare mass and the electric field, the color field should also be combined into the physical mass.
Thus, we need to evaluate the effective mass of the color field.
This goal has been realized in the perturbative sector of the quantum chromodynamics (QCD),
however, this task has not been finished and it might be complicate in the nonperturbative sector.

On the other hand, instead of the physical mass of a quark, the constituent quark degree of freedom (d.o.f) was introduced into a hadron system, which is dominated by the nonperturbative sector of QCD. Then, some interesting question arise. For example,
why is it so different between the pion defined as two quark system and the proton defined as three quark system?
Why is it so different between the masses of constituent quarks in a pion and the ones in a proton,
and, does this depend on the evaluation to the effective masse of the color field in the nonperturbative sector?

So, if we can evaluate the effective mass of the color field, either quantitatively or qualitatively,
then we might get some clues on the masses of constituent quarks,
and we might also get some information on the mass spectrum of hadrons with multiple quarks.
In this work, we will evaluate the effective mass of the color field in a classical field picture.

The remainder of this paper is organized as follows.
In Section \ref{Physical-dof}, we review the introduction of current quarks and give a definition on the physical d.o.f by going back to classical field picture from quantum field picture.
In Section \ref{Mass-spectrum}, we will try to give a hypothesis on the mass spectrum of compact multi-quark hadrons and glueballs, and try to explain why the compact $N$-quark hadrons still disappear so far.
In Section \ref{Constituent-quarks}, we will try to give some new understanding to the constituent quarks.
Finally, the conclusions
are given in Section \ref{conclusions}.

\section{Mass of Physical d.o.f: Back to Classical Field Picture from Quantum Field Picture}\label{Physical-dof}

Besides of the mass renormalization,
we will show, the physical dynamical quantum vacuum state will be related to the classical field.
The crucial reason is that, the high excited state $|N \hbar \omega\rangle$ with energy $N \hbar \omega$ of a simple harmonic oscillator system is equivalent to the Fock state of an $N$-particle system (or, the direct product of each single particle state $\hbar \omega$  with energy $\hbar \omega$);
and, like the laser,
by statistics on the probability amplitude of single particle,
the energy density distribution (not the normalized state vector) of an $N$-particle system is equivalent to the field strength  of a classical field.

For a simple harmonic oscillator system, the Hamiltonian is defined as\cite{Peskin}
\ba
H = \sum_{\bm k}  \omega_{\bm k} \left(a^\dag_{\bm k} a_{\bm k}  + \frac{1}{2}  \right);
\ea
and, the ground state (vacuum state) $| 0 \rangle $ and the single particle state $| 1\rangle $, the creation operator $a^\dag_{\bm k}$  and the annihilation operator $a_{\bm k}$, are defined as
\ba
| 0 \rangle = \prod_{\bm k} | 0 \rangle_{\bm k} ,\,\langle 0 | 0 \rangle = 1 ,\,\\
a_{\bm k} | 0 \rangle = 0, \, a^\dag_{\bm k} | 0 \rangle = | 1 \rangle,\,a_{\bm k} | 1 \rangle =| 0 \rangle .
\ea
The energy of vacuum state (the zero-point energy) is
\ba
E = \langle 0 | H |0 \rangle = \sum_{\bm k} \frac{1}{2}  \omega_{\bm k} = \infty ;
\ea
and, to avoid the infinity, a renormalized Hamiltonian is defined as
\ba
H \rightarrow H_{R} \equiv H -  \langle 0 | H |0 \rangle =   \sum_{\bm k}  \omega_{\bm k}  a^\dag_{\bm k} a_{\bm k} ,
\ea
with a ``single current quark" state $ |1_{m}   \rangle$ with the mass $m$ as the eigenfunction of $H_{R}$ defined as
\ba
H_{R} |1_{m}   \rangle\equiv  \omega_{m} | 1_{m}  \rangle ,\,
\omega_{m} = \sqrt{{\bm k}^2 +m^2}.
\label{define-currentquark-state}
\ea

However, by recalling the bare mass,
the energy of a free particle should be just infinity, and the infinity vacuum expectation value (VEV) should not be subtracted roughly; on the other hand, the eigen-equation of $H_{R}$ in (\ref{define-currentquark-state}) is not a real Schrodinger equation and can not include the full information of the system.
Indeed, for a full Hamiltonian $H$, the vacuum state $|0 \rangle$ and the ``single real quark" state $|1_{M_0}\rangle = a^\dag  | 0\rangle$ with the mass $M_0$  should be
\ba
H |0 \rangle &=& E_0 | 0\rangle,\\
H |1_{M_0}  \rangle&=&E_1 | 1_{M_0} \rangle ,\, \\
=(H_0 +  H_R) |1_{M_0}  \rangle &=& (E_0 + \omega   ) | 1_{M_0} \rangle ;
\ea
then, for a state $|1_{m}\rangle$ defined in (\ref{define-currentquark-state}),
there will be
\ba
|1_{m}   \rangle &=& a_0|0 \rangle + a_1 |1_{M_0}\rangle+ a_2 |2_{M_0}\rangle +...., \,
\label{current-quark-is-mixed-state} \\
\Rightarrow
\omega_{m} &\neq&  \omega, \, 0< \omega_{m} < +\infty,
\ea
or, inversely, from (\ref{current-quark-is-mixed-state}) there will be
\ba
|1_{M_0}  \rangle =  b_1 |1_{m} \rangle + b_0| 0 \rangle + b_2 |2_{M_0}\rangle +...;\, \label{physical-quark-is-mixed-state-0}
\ea
where
the coefficients $a_i$ and $b_i$ might be functions of many parameters (e.g., the energy scale $\mu$).
That means, the current quark state $  |1_{m}   \rangle$ should be a mixed state of the ``single real quark" state $ | 1_{M_0} \rangle$ and other higher energy level states;
and, in the perturbative sector, there should be $a_0 \simeq 0$ and $a_1 \simeq 1$ thus $  |1_{m} \rangle   \simeq  |1_{M_0}   \rangle $.\\

When the interaction $H_{int}$ is included, there will be the eigenvector of the total Hamiltonian $H$,
\ba
|1_{M}  \rangle &=& \alpha^\dag  | \Omega\rangle
=  c_1 |1_{m} \rangle + c_0| \Omega \rangle + c_2 |2_{M}\rangle +..., \,\non\\
\mbox{with}\quad &&
\alpha \neq a,\,
| \Omega \rangle \neq | 0 \rangle,\,
  \label{physical-quark-is-mixed-state}
\ea
where $| \Omega \rangle  $ is the new vacuum state,
the coefficients $c_i$ could be functions of many parameters (e.g., the energy scale $\mu$),
and $\alpha^\dag$ can be seemed as a Bogliubov transform of $a^\dag$;
we will call $|1_{M}  \rangle$ as the  ``single physical quark" state with the mass $M$.
From (\ref{physical-quark-is-mixed-state}), it means that,
$|1_{M}  \rangle$
should be a mixed state of the current quark state $  |1_{m}   \rangle$  and other multi-particle states;
and the mass relation will be
\ba
M &=&  |c_1|^2  m + | c_0|^2 m_\Omega +  \bar{M}  \non\\
\xrightarrow{\scriptsize \mbox{(ren.)}}&&  |c_1|^2  m +  \bar{M} . \, \label{mass-of-physical-quark-from-mixed-state}
\ea
In (\ref{mass-of-physical-quark-from-mixed-state}),
$m_\Omega \simeq\infty$ is the effective mass of $| \Omega \rangle$ state and it will be counteracted by
the ``infinite part" of
the color field surrounding the quark (by omitting the electric field), denoted by ${\bm E}_{c}^{(\infty)}$;
besides, $\bar{M}$ is the effective mass of a superposition state $|\bar{M}  \rangle \equiv c_2 |2_{M}\rangle +...$, and it will be corresponding to the ``finite part" of
the color field surrounding the quark, denoted by ${\bm E}_{c}^{(\bar{M})}$.
As mentioned above,
the high excited state $ |N_{M}\rangle $ of an effective harmonic oscillator is equivalent to
the Fock         state                          of an $N$-particle system,
and, like the laser, the energy $\bar{M}$ (not the normalized state vector) of an $N$-particle system is equivalent to the field strength of a classical field ${\bm E}_{c}^{(\bar{M})}$ .
In a word,
the mass of physical state $|1_{M}\rangle$ is the combination of the mass of current quark state  $ |1_{m}   \rangle$
and the mass of an effective classical field ${\bm E}_{c}^{(\bar{M})}$,
or,
the mass correction to a system of multiple quarks is
from the potential energy (interaction ) $H_{int}$,
which is just the energy of color field!

We should point out that, the ``single physical quark" state $|1_{M}\rangle$ (eigenstates of the full Hamiltonian $H$) is indeed a
``quasi-particle" state
rather than a real
``particle" state, since it cannot exist due to the confine character of QCD.
Instead, the ``single current quark" state $ |1_{m} \rangle$ is seemed as physical excitation quanta in QCD in the perturbation sense,
while the state $|1_{M}\rangle$ is traditionally called ``dressed quark" since the mass $\bar{M}$ could be seemed as a non-perturbative self-energy correction on a current quark, or
called
``constituent quark" \cite{HQ} which will be discussed in Sect.\ref{Constituent-quarks}.

Our model is different with the decomposition picture
in Ref.~\cite{BKG-field}.
In Ref.~\cite{BKG-field}, the field operators of quarks and gluons are redefined to the combination of
a background field part and a quantum field part and then a class of new Feynman rules of quarks and gluons were given,
that means,
the field theory is still constructed in a quantized scheme.

\section{\label{Mass-spectrum}Mass spectrum of compact multi-quark hadrons and glueballs}

\subsection{\label{model-build} A bag-tube oscillation model}

People have made lots of attempts to interpret the mass spectrum of compact $N$-quark hadrons
in many methods \cite{report2007}\cite{review1981}\cite{CQM}, such as
the lattice quantum chromodynamics (LQCD) \cite{LQCD},
QCD sum rules \cite{sum-rules},
and the constituent quark models,
et al.
Most of the constituent quark models are nonrelativistic,
that means, only the constituent quark d.o.fs
existed. In these nonrelativistic potential models,
due to the character of QCD,
one leading part of the potential is always constructed to a confining form, such as:
a volume-dependent form (e.g., the quark-bag model \cite{bag-model}),
a harmonic oscillator form (e.g., the Isgur-Karl model \cite{CQM}\cite{Isgur-Karl-model}
or Skyrme model in large $N_c$ limit \cite{large-Nc}),
a linear form
(e.g., the Cornell model \cite{Cornell}),
et al;
and the other leading part of the potential is always constructed to a Coulomb form or a Yukawa form;
in addition, as the next-to-leading part, some hyperfine potential terms
will be constructed
as well to generate the complicate hadronic mass spectrum, such as
the Regge trajectory terms (orbital and spin interactions) \cite{Regge-term},
the spin-flavor coupling terms (the G$\ddot{u}$rsey-Radicati terms)\cite{Spin-flavor-part},
and so on.
As converting
to the language of field theory, e.g., in the QCD framework,
a linear form confinement potential will arise in the nonperturbative regime of QCD, which
can be derived from the lattice QCD (LQCD)
calculation in the heavy quark limit\cite{LQCD-fit-potential};
some logarithmical form potential \cite{log-potential}
and
the Coumlomb form potential due to one gluon exchange processes will arise in the perturbative regime of QCD \cite{BSE}.
Or, instead the Coulomb form, a Yukawa form potential from one Goldstone boson exchange processes in chiral perturbative theory (ChPT)  will arise \cite{CHPT+DiracEq}\cite{chiral-constituent-quark-model}.

\begin{figure}[!htbp]
\centering \includegraphics[scale=0.5]{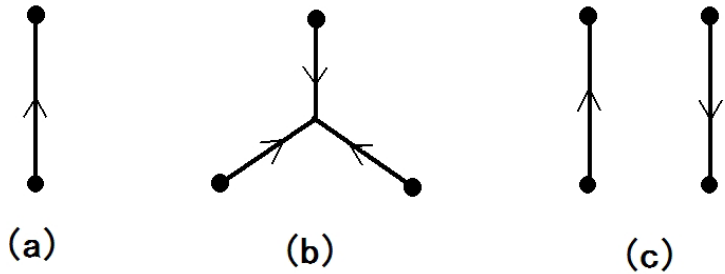}
\caption{In the ``flux-tube model" \cite{flux-tube-model-0}, flux tube configurations for (a) mesons, (b) baryons and (c) 4-quark hadrons, respectively;
particularly, (a) a flux tube (or "flux link" on the lattice) is a directed element (or "string");
(b) three units of flux all directed toward (away from) a ``junction" can annihilate (be created) there. }
\label{flux-tube-model}
\end{figure}

To understand the nonperturbative effect of QCD, especially the linear potential from LQCD, some pseudo-particle d.o.fs are proposed,
since gluon d.o.fs are not good ones in the nonperturbative regime.
On one hand, motivated by the linear Regge trajectories, mesons are treated as strings;
on the other hand, based on the QCD picture in strong-coupling regime, as what are shown in LQCD,
it is found that
a class of collective string-like flux tubes (or ``flux links" on the lattice) generated from the gluon condensations
could be seemed as good natural d.o.fs (for the reason, e.g., in lattice QCD, for the color-electric field $\bm E$, $|\bm E|^2$ has a definite nonzero eigenvalue).
So, a type of so-called flux-tube model (or called string-like model, collective model or hypercentral model)
is established \cite{flux-tube-model-0}\cite{flux-tube-model}.
However, that does not mean flux tubes are really physical objects; instead, they are just pseudo-particle d.o.fs.
Here we just list some details on the flux-tube model which will be useful to introduce our own model, as below: \\
\indent (1) flux tubes are directed, with a quark or antiquark acting as a unit source (sink) of flux;\\
\indent (2) the configurations of mesons, baryons and  $4$-quark hadron states are shown in Fig.~\ref{flux-tube-model}-(a), \ref{flux-tube-model}-(b) and \ref{flux-tube-model}-(c), respectively;
for baryons in Fig.~\ref{flux-tube-model}-(b), there is a junction of 3 flux tubes;
for $4$-quark hadrons or multi-quark hadrons in Fig.~\ref{flux-tube-model}-(c), there exists mixing between different topological configurations;
the dynamics of multi-quark states are highly not like mesons and baryons, and the system might not be completely confined \cite{flux-tube-model-0}\cite{no-Nq};\\
\indent (3) dynamics: hadrons decay via flux tube breaking, due to quark pair creation;  \\
\indent (4) for color-singlet hadrons, the residual force (van der Waals-type force) will be suppressed to a short-range one,
due to the confinement or color screening;  \\
\indent (5) when all the flux tubes in a hadron are ``frozen" (on the so-called adiabatic surface),
the hadron is a purely quark-flavored state with definite quantum numbers,
and with energy $E_0 = b_0 \sum_{n=1}^{N} R_n $, where $N$ is the number of flux tubes in the hadron;
when there are excitation modes (phonon) with energy $\{\omega_n, \,n=1,2,...,n_{max}\}$ on the flux tubes, the hadron will become a hybrid,
with energy $E_0 +\sum_n  \omega_n$;
that means, the energy of a hadron includes a linear potential energy from the string-like tensional energy between the static quarks as leading part, with a kind of harmonic vibrational energy as perturbation; \\
\indent (6) glueball states are purely consisted with flux tube loops without quarks; etc.

\begin{figure}[!htbp]
\centering \includegraphics[scale=0.7]{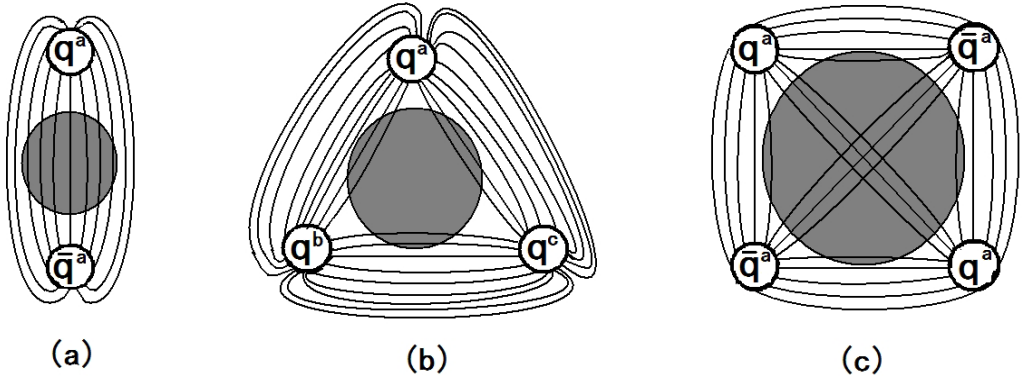}
\caption{In our ``bag-tube oscillation model", the tube configurations of color fields for (a) mesons, (b) baryons and (c) 4-quark hadrons, respectively, with the superscripts $a,b,c$ denoting the color indices.}
\label{tube-volume-0}
\end{figure}

Now,
we will propose our own model. Our model can be seemed as a kind of combination of quark-bag model\cite{bag-model} and flux tube model,
and we call our model ``{\bf bag-tube model}" or ``{\bf bag-tube oscillation model}".
Here we only list some tentative definitions
for our model, as below: \\
\indent (1) we define that, the space configuration of a hadron is in an oscillation between a bag-like one and a tube-like one, i.e.,
color fields induced by the color charges (quarks, antiquarks, or gluons) in a hadron are confined in a bag when charges are in short distances,
while the color fields are confined in the tube volumes when charges are in long distances;
we want to stress that, each color charge does not only act as a unit source of color fields in all the tubes linked with itself,
it will also
influence the strength of
color fields in all the other tubes not linked with itself
in the whole hadron, as shown in formula  (\ref{density-on-N2}) in Sect. \ref{mass-on-N4};\\
\indent (2)
we propose that, the flux tubes should exist between each two sources, because ``a quark could not know which one(s) it would choose to be teamed with" (introducing the topological mixing in Fig.~\ref{flux-tube-model}-(c) is just a solution to this ``problem");
in our picture,  color charges are only responsible for producing color fields, and, the confined space configuration and quantum numbers of a hadron should be fixed until (or, by) a projection onto the Hilbert space;
so,
the configurations of mesons, baryons and  $4$-quark hadron states are shown in Fig.~\ref{tube-volume-0}-(a), \ref{tube-volume-0}-(b) and \ref{tube-volume-0}-(c), respectively, e.g.:
for baryons in Fig.~\ref{tube-volume-0}-(b), the configuration of tubes is a ``$\triangle$-type" instead of the ``Y-type" in Fig.~\ref{flux-tube-model}-(b), and for $4$-quark hadrons,
the configuration of tubes is the one in Fig.~\ref{tube-volume-0}-(c) instead of the one in Fig.~\ref{flux-tube-model}-(c);
we ignore the direction of flux tube at the first sight;
the gray bags in Fig.~\ref{tube-volume-0}-(a), (b) and (c) are the quark-bag configurations;
all the multi-quark systems could be confined, but not all of their lives would be long, see Sect. \ref{disappear-due-to-heavy};\\
\indent (3)
dynamics: in flux tube models reviewed above, hadrons decay via flux tube breaking due to quark pair creation, however, we should ask:
where will a string break down (or, where is the most solid/firm/stable place of a sting), at the high energy density region or low energy density region (or, at the weak force region or strong force region)? where is the high energy density region (or, where is the strong force region), the region of quarks colletion or the region of quarks diffusion?
in our model, we skip the string picture;
instead, in our picture,
we define, a hadron can decay via emitting color-singlet quark clusters (hadrons),
which are generated by the recombination of color charges
when a hadron turns
to the bag configuration phase during the oscillation;\\
\indent (4)  for color-singlet hadrons, the residual force (van der Waals-type force) will be suppressed to a short-range one, due to the confinement or color screening;  \\
\indent (5) we define that,
the energy of a compact $N$-quark hadron has a leading part of
\ba
E_0= \rho^c_N V_N = \rho^c_N  \cdot n  V_2 ,\, \label{bag-tube-model}
\ea
where $n$ is the number of tubes in the hadron,
$V_2$ is the equal constant volume value of all the tubes in a hadron, see (\ref{volume=CN2}) in Sect. \ref{mass-on-N4},
that means, the shape of the tubes can vary but its volume will not change; \\
\indent (6)
in glueball states, it is also the gluonic color charges that act as sources of color fields,  like in the multi-quark states, see Sect. \ref{Glueballs}; etc.

\subsection{\label{mass-on-N4}A dependence of mass on $N^4$}

Like the Coulomb fields,
for an effective classical field ${\bm E}_{c}^{(\bar{M})}$ induced by a source with color charge quantum number $Q^{c}$,
it is reasonable to
measure the field energy density as
\ba
\rho^{c} \sim  |{\bm E}_{c}^{(\bar{M})}|^2 \sim |Q^{c}|^2; \label{density-on-N2}
\ea
so,
for an $N$-quark hadron state with definite flavor and color wavefunction
\ba
hadron = \hat{\mathcal O}_{ir}\left[   q_{f_1}^{c_1}  q_{f_2}^{c_2} \ldots  q_{f_N}^{c_N} \right],
\ea
where the notation $\hat{\mathcal O}_{ir}$  denotes an operation to pick out one eigenstate in the irreducible representation
of the direct product group of flavor-symmetry and color-symmetry,
it is reasonable to
measure the field energy density as
\ba
\rho^{c}_N \sim  |N \cdot {\bm E}_{c}^{(\bar{M})}|^2 \sim |N Q^{c}|^2 = N^2 | Q^{c}|^2 = N^2 \rho^{c}_1,
\ea
where we take $\rho^{c}_1$ to denote the color field energy density stimulated by one ``single physical quark".

To evaluate the ``effective volume" of the color field configuration in a hadron, we have recalled the tube configuration \cite{Peskin} of a color field between two color sources, see Fig. \ref{tube-volume-0},
and it will be reasonable to
measure the effective volume as
\ba
V_N \sim C_{N}^{2} V_2 \sim N (N-1) V_2 \xrightarrow{\scriptsize \mbox{$N\gg 1$}} N^2 V_2   , \,
N \geq 2,  \label{volume=CN2}
\ea
where $V_2$ is taken to denote the volume of one single tube formed by 2 quarks,
and $C_{N}^{2} \equiv \frac{N!}{(N-2)!2!}= \frac{N (N-1)}{2}$ is the binomial coefficient.
Here we should not confuse the color field with the electromagnetic field, for example,
the r.m.s electric charge radius of pion (about $0.659 \,fm$) and proton (about $0.8409 \,fm$)\cite{PDG-2020} will give a ratio of the electromagnetic volume $0.659^3 /0.8409^3  \simeq 1/ 2.078 $, while the ratio of color volume is about $1/3$ from (\ref{volume=CN2}).

Thus, the total energy of the color field in a compact $N$-quark hadron will roughly have a dependence on $N^4$
in the increasement of $N$, as
\ba
\bar{M}_N \equiv \rho^{c}_N  V_N
&\sim& N^2\cdot  N (N-1) \bar{M}_0
\label{mass=rho-dot-V}\\
&\xrightarrow{\scriptsize \mbox{$N\gg 1$}} &  N^4 \bar{M}_0  ,\,
N \geq 2,\,
\bar{M}_0 \equiv  \rho^{c}_1 V_2 .
\ea
Coincidentally,
if we formally treat the Hamiltonian operator as $\hat{H} \sim |E\rangle \langle E|$ for an energy eigenstate $|E\rangle$ of a ``single physical quark", then,
after mapping the state $|E\rangle$ to a classical field strength ${\bm E}$, the energy of a ``single physical quark" will be
$E_1 = \langle E| \hat{H} |E\rangle  \sim  \langle E|\cdot |E\rangle \cdot \langle E| \cdot |E\rangle \sim | {\bm E}|^4 $,
and the energy of an $N$-quark hadron will become
$E_N \sim N^4 | {\bm E}|^4$.

In combination with the current quark mass $m$, as shown in (\ref{mass-of-physical-quark-from-mixed-state}),
the total mass of a compact $N$-quark hadron will be
\ba
M_N &\sim&  ( m_1 + m_2 +\ldots +m_N ) +  \bar{M}_N   \non\\
&\sim&   ( m_1 + m_2 +\ldots+ m_N ) +  N^2\cdot  N (N-1) \cdot   \bar{M}_0 ; \label{mass-rule-JL0}
\ea
moreover, to include the information of Regge trajectory on the angular momentum $L$ and the total spin $J$ of a hadron, we modify the total mass above to be
a hypothesis, as
\ba
M^{JL}_{N}  =    ( m_1 + m_2 +\ldots + m_N ) +  N^2 \cdot  N (N-1) \cdot a_{N}^{JL} + b_{N}^{JL}, \label{mass-rule}
\ea
where $a_{N}^{JL}$ and $b_{N}^{JL}$ are dimensionful constants for definite $\{N,J,L\}$ configuration.
The coefficient $|c_1|^2$ in (\ref{mass-of-physical-quark-from-mixed-state}) has been absorbed into $m_i$ in (\ref{mass-rule-JL0},\ref{mass-rule}).

\subsection{Current quark mass $m_i$ defined in a perturbation sense}

Now we will concentrate on the values of current quark masses $m_i$ in
(\ref{mass-rule}).
What is $m_i$?
From (\ref{define-currentquark-state}), $m_i$ is originally defined as the mass of free current quark,
but, as said in the end of Sect. \ref{Physical-dof}, the current quark is seemed as physical d.o.f in QCD in the perturbation sense,
so $m_i$ can be also seemed as
the pole mass of the current quark defined by the pole position in the full propagator in the perturbation sense in QCD \cite{PDG-2020}.
In the perturbation sense, there is a relation between the pole mass $m_f$ of current quark with flavor $f$ and the $\overline{MS}$ ``running¡± mass $\overline{m}_f(\mu)$ in the perturbation sense \cite{PDG-2020}\cite{RGE},
and $\overline{m}_f(\mu)$ is conventionally defined at a scale $\mu \gg \Lambda_{\chi}$,
where $\Lambda_{\chi} \sim 1\,GeV$  is the non-perturbative scale of dynamical chiral symmetry breaking,
for example,
$\overline{m}_s(2 GeV)\sim 100 \,MeV$,
$\overline{m}_c(\overline{m}_c)=1270 \,MeV$,
$\overline{m}_b(\overline{m}_b)=4180 \,MeV$.

However,
here the challenge is an inverse problem, that is,
how can we determine the energy scale $\mu$, or furthermore, how can we determine $m_i(\mu)$, $a_{N}^{JL}(\mu)$,  $b_{N}^{JL}(\mu)$ and $M^{JL}_{N}(\mu)$
(since generally the variables $m_i$, $a_{N}^{JL}$,  $b_{N}^{JL}$ and $M^{JL}_{N}$ in (\ref{mass-rule})
are all energy scale dependent)?
Moreover,
is it possible for $\mu \simeq \Lambda_{\chi}$ (located in the non-perturbative region) rather than $\mu \gg \Lambda_{\chi}$ (located in the perturbative region)?

On the other hand, how to examine our ``$N^4$ mass rule" hypothesis in (\ref{mass-rule})?
One method is to check the reasonability of derived values of $m_i$ after accepting the ``$N^4$ mass rule",
that is, the validity of the ``$N^4$ mass rule" can be indirectly confirmed
by the reasonability of derived values of $m_i$.
In detail,
after fixing the values of $m_{u,d}$, $a_{N}^{JL}$ and $b_{N}^{JL}$,
then with Eq. (\ref{mass-rule}), we will
compute $m_{s,c,b}$ for hadrons,
and we can check whether the $m_{s,c,b}$ values follow appropriate requirements or not.
We will choose the values below (only for $L=0$):
\ba
&&m_q  = m_u = m_d = 10 \,MeV ,\,(q = u,d),\label{mu-and-md}\\
&&b^{J,L=0}_{N}\simeq S(S+1)\cdot\frac{305}{C_N^2}    \,MeV,\,(S=J=0,\frac{1}{2},1,\frac{3}{2};\, N=2,3), \label{bJLN-coefficient}\\
&&a^{J=0,L=0}_{N=2}=a^{J=\frac{1}{2},L=0}_{N=3}=a^{J=1,L=0}_{N=2}=a^{J=\frac{3}{2},L=0}_{N=3} \simeq 15 \,MeV;\,\label{aJLN-coefficient}
\ea
here Eq. (\ref{mu-and-md}) is motivated by the mass running effect, i.e., the value of $m_q$ in hadrons should be larger than the current $d$ quark mass value $4 \,MeV$ at $\mu=2GeV$ \cite{PDG-2020};
Eq. (\ref{bJLN-coefficient}) is  motivated by
the mass difference between $\pi(140)$ and $\rho(775)$ and the  mass difference between $p(940)$ and $\Delta(1232)$;
Eq. (\ref{aJLN-coefficient}) is derived by inserting (\ref{mu-and-md},\ref{bJLN-coefficient}) and the mass of $\pi(140)$ into Eq. (\ref{mass-rule});
and we should note that, here we only define values for the $L=0$ case, so we can ignore the $\sqrt{L}$ type terms of Regge Trajectories.
The results of $m_{s,c,b}$ are listed in Table \ref{table-s-quark}, \ref{table-c-quark} \ref{table-b-quark}
respectively.

\begin{table}[!htbp]
\caption{\label{table-s-quark} Values of current quark mass $m_s$ computed with $ m_u = m_d = 0$.
}
%\begin{ruledtabular}
\begin{tabular}{|c||c|c|c||c|c|c|}
\hline
flavor     & mass (MeV)              &    $m_s/MeV$    & prediction &mass (MeV)       &      $m_s/MeV$      & prediction  \\
\hline
$q\bar{q}$ &$\pi(140)$               &                    &         &$\rho,\omega (775)$ &                  &     \\
$q\bar{s}$ &$K(494)$                 &  $364$           &         &$K^{\ast}(892)$     &     $152$           &     \\
$s\bar{s}$ &$\eta(550)$              &  $?$            & mixing?  &$\phi(1020)$        &    $145$              &     \\
\hline\hline
$qqq$ & $p,n(940)$                  &                   &         &$\Delta(1232)$      &           &       \\
\hline
$qqs$ & $\Sigma(1200),\Lambda(1116)$ & $293,209$       &         &$\Sigma^{\ast}(1385)$&      $173$                   &        \\
$qss$ & $\Xi(1320)$                  & $211$            &         &$\Xi^{\ast}(1530)$   &     $164$           &        \\
$sss$ &     $--$                     & $--$            &    $--$   &$\Omega(1672)$       &    $160$              &        \\
\hline
\end{tabular}
%\end{ruledtabular}
\end{table}

\begin{table}[!htbp]
\caption{\label{table-c-quark} Values of current quark mass $m_c$ computed from  Eq. (\ref{mass-rule}) with $ m_u = m_d = 0$, $m_s=\overline{m}_s(2 GeV)\sim 100 \,MeV$.
Each hadron mass range (in the unit $MeV$) in the ``prediction" columns denoted as ``$(\,,\,)$", is corresponding to a $m_c$ range put by hand and denoted with a colon as ``$:(\,,\,)$". }
%\begin{ruledtabular}
\begin{tabular}{|c||c|c|c||c|c|c|}
\hline
flavor     & mass (MeV)              &    $m_c/MeV$    & prediction &mass (MeV)       &      $m_c/MeV$      & prediction  \\
\hline
$c\bar{q}$ &$D(1870)$                &  $1740$          &         &$D^{\ast}(2010)$    &    $1270$       &     \\
$c\bar{s}$ &$D_s(1968)$              &  $1748$           &         &$D^{\ast}_s(2112)$  &   $1282$           &     \\
$c\bar{c}$ &$\eta_c(1S)(2980)$       &  $1430$           &         &$J/\psi(3097)$      &  $1183$           &     \\
\hline\hline
$qqc$ & $\Sigma_c(2455),\Lambda_c(2286)$ & $1548,1379$   &         & $\Sigma^{\ast}_c(2520)$ & $1308$ &        \\
\hline
$qsc$ & $\Xi_{c}(2468)$, $\Xi'_{c}(2578)$& $1471,1581$   &         & $\Xi^{\ast}_{c}(2645)$  & $1343$  &        \\
$qcc$ & $\Xi_{cc}(3621?)$                &  $1362?$      & unknown &$\Xi^{\ast}_{cc}(3621?)$ & $1209?$  & unknown  \\
\hline
$ssc$ &     $--$                         &  $--$  &  $--$    & $\Omega_c(2770)$        & $1378$        &        \\
$scc$ &     $--$                         &  $--$  &  $--$    & $\Omega_{cc}(?)$        & $:(1270,1378)$ & $(3831,4047)$  \\
$ccc$ &     $--$                         &  $--$  &  $--$    & $\Omega_{ccc}(?)$       & $:(1270,1378)$ & $(5001,5325)$ \\
\hline
\end{tabular}
%\end{ruledtabular}
\end{table}

\begin{table}[!htbp]
\caption{\label{table-b-quark} Values of current quark mass $m_b$ computed from  Eq. (\ref{mass-rule}) with $ m_u = m_d = 0$, $m_s=\overline{m}_s(2 GeV)\sim 100 \,MeV$ and
$m_c=\overline{m}_c(\overline{m}_c)=1270 \,MeV$.
Each hadron mass range (in the unit $MeV$) in the ``prediction" columns denoted as ``$(\,,\,)$", is corresponding to a $m_b$ range put by hand and denoted with a colon as ``$:(\,,\,)$". }
%\begin{ruledtabular}
\begin{tabular}{|c||c|c|c||c|c|c|}
\hline
flavor     & mass (MeV)              &    $m_b/MeV$    & prediction &mass (MeV)       &      $m_b/MeV$      & prediction  \\
\hline
$b\bar{q}$ &$B(5279)$                &  $5149$        &         &$B^{\ast}(5325)$    &      $4585$       &     \\
$b\bar{s}$ &$B_s(5366)$              &  $5146$        &         &$B^{\ast}_s(5413)$  & $4583$              &     \\
$b\bar{c}$ &$B_c(6275)$              &  $4885$        &         &$B^{\ast}_c(?)$     &                   &     \\
$b\bar{b}$ &$\eta_b(1S)(9300)$       &  $4590$        &         &$\Upsilon(1S)(9460)$& $4365$              &     \\
\hline
\hline
$qqb$ & $\Sigma_b(5810),\Lambda_b(5620)$ &  $4903,4713$  &      & $\Sigma^{\ast}_b(5830)$ & $4618$ &        \\
\hline
$qsb$ & $\Xi_b(5797)$,$\Xi'_b(5935)$ &  $4800,4938$  &      & $\Xi^{\ast}_b(5955)$    & $4653$  &        \\
$qcb$ & $\Xi_{cb}(?)$     & $:(4180,4800)$   & $(6346,6966)$     & $\Xi^{\ast}_{cb}(?)$    & $:(4180,4653)$ & $(6651,7124)$       \\
$qbb$ & $\Xi_{bb}(?)$     & $:(4180,4800)$   & $(9256,10496)$    & $\Xi^{\ast}_{bb}(?)$    & $:(4180,4653)$ & $(9561,10507)$      \\
\hline
$ssb$ &    $--$                         &  $--$  &  $--$     & $\Omega_b(6046)$        & $4654$ &        \\
$scb$ &    $--$                         &  $--$  &  $--$     & $\Omega_{cb}(?)$        & $:(4180,4654)$ & $(6741,7215)$ \\
$sbb$ &    $--$                         &  $--$  &  $--$     & $\Omega_{bb}(?)$        & $:(4180,4654)$ & $(9651,10599)$  \\
$ccb$ &    $--$                         &  $--$  &  $--$     & $\Omega_{ccb}(?)$       & $:(4180,4654)$ & $(7911,8385)$  \\
$cbb$ &    $--$                         &  $--$  &  $--$     & $\Omega_{cbb}(?)$       & $:(4180,4654)$ & $(10821,11769)$  \\
$bbb$ &    $--$                         &  $--$  &  $--$     & $\Omega_{bbb}(?)$       & $:(4180,4654)$ & $(13731,15153)$  \\
\hline
\end{tabular}
%\end{ruledtabular}
\end{table}

\begin{figure}[!htbp]
\centering \includegraphics[scale=1.0]{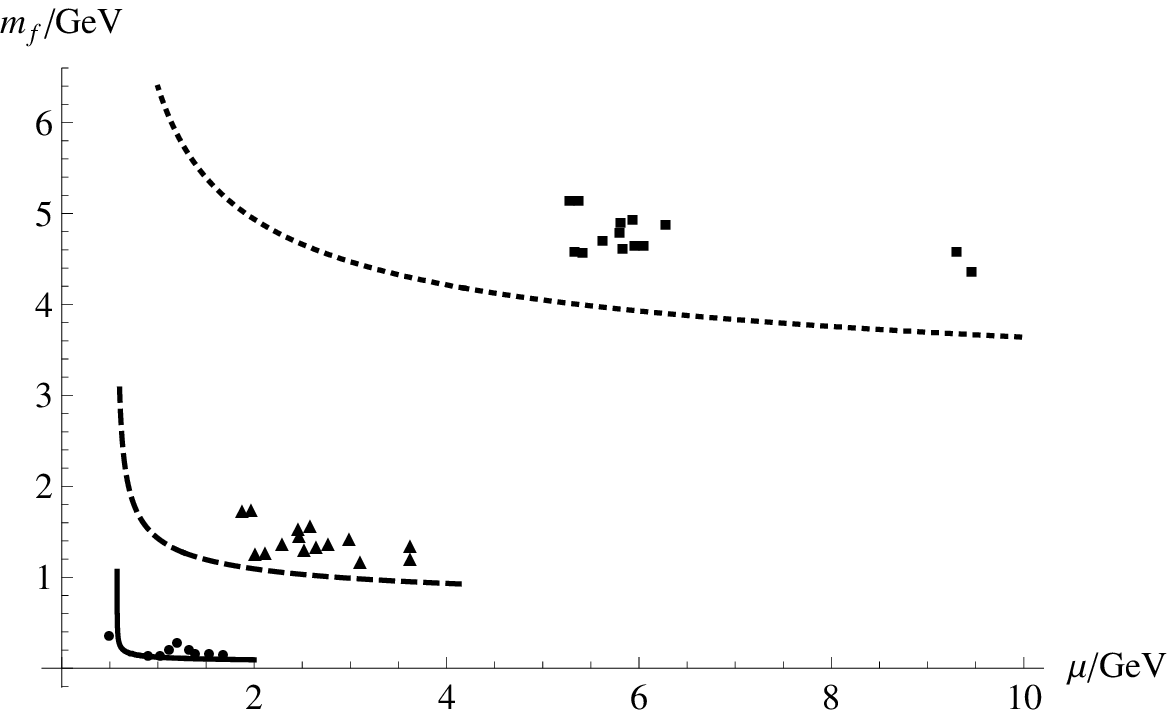}
\caption{The solid, dashed, dotted lines are respectively for dependence of the running mass $m_f$ of current quark with flavor $f=s,c,b$ on the energy scale $\mu$, which are naively derived in the perturbative approach within the $\overline{MS}$ scheme \cite{RGE};
the rounded, triangled, squared points are respectively for the current quark masses of $m_{s}$, $m_{c}$, $m_{b}$ extracted in a non-perturbatively approach via our model (\ref{mass-rule}), with the corresponding $\mu$ values naively set to the mass of each corresponding hadron only just for convenience.
%(***************************************************************)
For the $m_{b}$ line, we set $\alpha_s (m_b)= 0.223$ \cite{PDG-2020}
and $N_f = 5$
in the range of $\overline{m}_b(\overline{m}_b) < \mu < 10 \,GeV$,
with  $N_f = 4$ in $1 \,GeV < \mu <\overline{m}_b(\overline{m}_b)$;
%(***************************************************************)
for the $m_{c}$ line, we set $N_f = 4$ in $\overline{m}_c(\overline{m}_c) < \mu <\overline{m}_b(\overline{m}_b)$,
with $N_f = 3$ in $0.6   \,GeV < \mu <\overline{m}_c(\overline{m}_c)$;
%(***************************************************************)
for the $m_{s}$ line, we set $N_f = 4$ in $\overline{m}_c(\overline{m}_c) < \mu < 2\, GeV$,
%(************)
with $N_f = 3$ in $0.35\, GeV < \mu < \overline{m}_c(\overline{m}_c)$.
%(***************************************************************)
}
\label{RGE-mass-scb}
\end{figure}

In Fig.~\ref{RGE-mass-scb},
the results of the current quark masses $m_{s,c,b}$ extracted in a non-perturbative approach via our model (\ref{mass-rule})
are plotted as discrete points (rounded, triangled or squared, respectively),
while
the results
naively derived in the perturbative approach within the $\overline{MS}$ scheme \cite{RGE} are plotted as curves.
We should point out that,
the energy scale $\mu$ of each discrete point plotted in Fig.~\ref{RGE-mass-scb} is just naively set to the mass of each corresponding hadron for convenience, however, the typical momentum transfer between two quarks in the hadrons should not be so large;
it is shown that, at least,
the range of of $m_{s}$ (or $m_{c}$, $m_{b}$) represented by the discrete points can be qualitatively consistent with the curves.
So, it would be reasonable to interpret the points in Fig.~\ref{RGE-mass-scb}
as the corresponding results of running current quark masses in the non-perturbative region of the (full) QCD,
that means, our model (\ref{mass-rule}) is reasonable in some sense.

Besides, we want to discuss a special
invalid sector of our $N^4$ model, that is, the singlet $\eta(550)$ sector.
Mesons
can be embeded in the representations ${\bm 8}$ and ${\bm 1}$ of the $SU(3)$ group of flavor symmetry, as
\ba
{\bm 8}:\quad  \eta_8(\omega_8)&=&(u\bar{u}+d\bar{d}-2s\bar{s})/\sqrt{6},  \label{flavor-8-let}\\
{\bm 1}:\quad  \eta_0(\omega_0)&=&(u\bar{u}+d\bar{d}+s\bar{s})/\sqrt{3}, \label{flavor-1-let}
\ea
while the physical states (mass eigenstates) are always the (non-ideal) mixing states, such as: \\
the pseudoscalar ones (with the mixing angle $\theta_P  =  -24.5^{\circ}$)
\ba
\eta(550) &=& \eta_8 \cos\theta_P  -  \eta_0\sin\theta_P
\simeq    0.61 \cdot u\bar{u}+ 0.61 \cdot d\bar{d}-0.51 \cdot s\bar{s}    ,\\
\eta'(958)&=& \eta_8\sin\theta_P  +  \eta_0\cos\theta_P
\simeq 0.36 \cdot u\bar{u}+0.36 \cdot d\bar{d}+0.86 \cdot s\bar{s},\\
\Rightarrow
m_{\eta_8} &=&
620\,MeV,\,
m_{\eta_0}
=888\,MeV,
\ea
and the vector ones  (with the mixing angle $\theta_V  =  36.5^{\circ}$)
\ba
\phi(1020) &=& \omega_8\cos\theta_V  -  \omega_0\sin\theta_V
\simeq  -0.01 \cdot u\bar{u}  -0.01 \cdot d\bar{d}-0.99 \cdot s\bar{s} ,\\
\omega(782) &=& \omega_8\sin\theta_V  +  \omega_0\cos\theta_V
\simeq 0.70 \cdot u\bar{u}+ 0.70 \cdot  d\bar{d} -0.02 \cdot s\bar{s}  ,\\
\Rightarrow
m_{\omega_8}&=&
936\,MeV,\,
m_{\omega_0}
=866\,MeV.
\ea
From (\ref{flavor-8-let}) and (\ref{flavor-1-let}), there would be $m_{\eta_8} > m_{\eta_0} $ and $m_{\omega_8} > m_{\omega_0} $ due to the larger proportion of $s$ quark in $\eta_8(\omega_8)$ than the one in $\eta_0(\omega_0)$,
however, there exsit $m_{\eta_8} < m_{\eta_0} $. What does this imply? One of the reasonable possibility is
that, there exist heavier partner particles with the same quantum numbers as of the singlet $\eta_0(q\bar{q})$, e.g., the multi-quark states $(q\bar{q}q\bar{q})$ or the hybrid states $(q\bar{q}g\bar{g})$,
which could enlarge the mass of $m_{\eta_0}$ due to the mixing.
So, the $N^4$-rule in our mass model (\ref{mass-rule}) would not hold well for the lighter hadrons $\eta_8(\omega_8)$ and $\eta_0(\omega_0)$, neither for $\eta(550)(\phi(1020))$ and $\eta'(958)(\omega(782))$.
So,
in Table \ref{table-s-quark},
we have not extracted the mass of current $s$ quark for $\eta(550)$.
Besides, we treat $\omega$ as pure $q\bar{q}$ state due to the very few proportion of $s$ quark.

\subsection{Match with results in chiral perturbative theory} \label{match-ChPT}

In chiral perturbative theory (ChPT),
the Goldstone bosons
will become pseudo-Goldstone bosons
with nonzero masses
due to the chiral symmetry breaking Lagrangian terms from small nonzero current quark mass $m_q$ ($m_q \ll \Lambda_{QCD}$) \cite{HQ},
\ba
{\mathcal L}_{m_q} &=& m_q \bar{q}{q},\, q=u,d,s,  \label{quark-mass-term}\\
\rightarrow
{\mathcal L}_{m_\chi} &=& v Tr(m_\chi^\dag \Sigma + m_\chi \Sigma^\dag),\, m_\chi\equiv m_q,  \label{chiral-symmetry-breaking-term}
\ea
where
$v \equiv \langle \Omega | \bar{q}(x) q(x)  |\Omega \rangle$
is the vacuum expectation value (VEV) of quark condensate,
and
$\Sigma = \exp[i 2 \mathbb{M}/f_\pi]  \sim \bar{q}^j_R(x) q^k_L(x)/v$ gives the local orientation of the quark condensate,
with $\mathbb{M}$ a $3\times 3$ hermitian matrix for mesons and $f_\pi$ the decay constant of pions.
For example,
the pions will have a mass dependence as
\ba
m_{\pi}^{2} \sim \frac{4 v}{f_\pi^2} m_\chi,  \label{mpion-on-mq}
\ea
that is, if we identify $m_\chi\equiv m_q$, then, $m_\pi$ is linear on the current quark mass $m_q$ and it will become zero in the chiral limit.
Thus, does that mean,
$m_\pi$ is linearly dependent on the quark mass $m_q$
and
$m_\pi$ is not dependent on the ``volume energy" in (\ref{bag-tube-model}) at all,
or, our model is wrong?
Solutions to this doubt can be listed below:

{\bf (1) One choice is to modify our model.}

For a meson, it can be massless in the chiral limit, provided its volume is zero;
this condition can be satisfied,
because a massless particle will move in speed of light, so its volume will automatically be zero.
For
a baryon, it can automatically avoid the massless case even in the chiral limit,
since there is always nonzero spin term $b^{JL}_N$ in our mass rule (\ref{mass-rule});
that is consistent with the case in ChPT.
That means, our model can still hold in the chiral limit.

Besides,
like the relation in (\ref{mpion-on-mq}) which is constructed to describe the restoration of chiral symmetry from breaking phase,
here we also need an
assumption
on the continuous transition from nonzero to zero for the volume of pion, $V_\pi$, or the common parameter $V_2$ in our model (by recalling $V_\pi = V_2$),
as
\ba
V_2 =V_\pi  \sim  m_\chi \equiv m_q , \quad \mbox{(only for $m_\chi \ll \Lambda_{QCD}$)} . \label{volume-on-mq}
\ea
However, we should stress that, $m_\chi\equiv m_q$ is introduced as just a parameter in (\ref{volume-on-mq}) rather than the real mass of quark,
and, once the parameter $m_\chi$ is fixed, the common volume parameter $V_2$ is independent on the flavors in a hadron.
Moreover,
like what people have done in ChPT, that is,
only for the $m_\chi\equiv m_q \ll \Lambda_{QCD}$ case, the chiral symmetry breaking effects can expressed in
an expansion on $m_q$ (or $\frac{m_q}{\Lambda_{QCD}}$) in (\ref{chiral-symmetry-breaking-term})
or in a mass relation in (\ref{mpion-on-mq}), or in our volume relation in (\ref{volume-on-mq});
otherwise, for large $m_q$, the chiral symmetry is explicitly broken and there could not perform an perturbative expansion on $\frac{m_q}{\Lambda_{QCD}}$
based on a symmetry theory any more,
or to say, there even could not exist a ChPT any more.

Although the real quark masses are actually not zero
and we can avoid the $m_\pi = 0$ case in our model,
nevertheless,
the continuous transition from nonzero to zero for the masses of Goldstone bosons
is very important to avoid criticisms for
a theory on approximate symmetry.

{\bf (2) The other choice is to modify ChPT.}

We would ask, does the Goldstone theorem still hold
even for composite particles
generated from the quark condensate due to a strong interaction?
Or, are the mesons really massless in the chiral limit?
Is it possible that, mesons are always massive, no matter whether quarks are massive or not?
Could we reinterpret the results in ChPT? For these questions, we are motivated by the three details below:

(i) In (\ref{quark-mass-term}) and (\ref{chiral-symmetry-breaking-term}),
although they are both the explicitly chiral symmetry breaking terms,
$m_\chi$ in (\ref{chiral-symmetry-breaking-term}) is not necessarily and certainly to be $m_q$ in (\ref{quark-mass-term})!
Otherwise, if $m_\chi$ is the quark mass $m_q$, one should answer, what is the value of corresponding energy scale $\mu$ to define this $m_q$?

(ii) Even in chiral symmetry reserved case, there can still generate meson mass terms, e.g., from the four quark coupling terms,
\ba
{\mathcal L}_{4q} \sim  \frac{\lambda}{v} \bar{f}_L f_L \bar{f}_R f_R  ,\, \lambda\equiv m_q . \label{4q-term}
\ea
In (\ref{4q-term}), if the dimensional parameter $\lambda$ is defined to $m_q$, i.e., $\lambda\equiv m_q$, then the meson can get a mass
$m_\pi^2 \sim \frac{v\lambda}{f_\pi^2}$;
so, is that just the underlying reason why the relation (\ref{mpion-on-mq}) holds so well?

(iii) To investigate the property of a meson in ChPT,
since mesons are generated from quark condensate,
one should consider both the VEV $v$ (including spontaneously vacuum symmetry breaking information) and mass parameter $m_q$ or $m_\chi$ (including the explicit chiral symmetry breaking information) at the same time rather than separately.

Here we want to give a new interpretation
to some results in the ChPT.
For example, at the quark level, if we take the assumptions for $v$ and $f_\pi$ as
\ba
\frac{v}{f_\pi}
&\equiv&\langle \Omega|  \frac{1}{f_\pi}\bar{q}(x) q(x) | \Omega  \rangle
\equiv \langle\frac{\varphi(x)}{f_\pi}\rangle
\sim  \frac{\varrho_g + \varrho_q}{(m_u + m_d)} \frac{1}{f_\pi},  \label{VEV-on-rho}\\
f_\pi^2 m_\pi &\sim& f_\pi \langle0|\bar{q}\gamma^0\gamma^5 q |\pi\rangle
\sim f_\pi  \cdot \Phi_{\bar{q}q}(x=0) \cdot {\mathcal M}_{\bar{q}q\rightarrow |0\rangle} \sim \frac{1}{V}, \label{fpion-on-V}
\ea
then we can get the mass relation
\ba
m_{\pi}
\sim (\varrho_g + \varrho_q)  V, \quad  \label{mpion-on-rhoV}
\ea
where the $\varrho_{g} V$ part can match with our hypothesis on effective energy $\bar{M}_N = \rho^{c}_N  V_N$ for gluon fields in (\ref{mass=rho-dot-V}).
Here in (\ref{VEV-on-rho}) we understand $v \equiv\langle \varphi(x) \rangle $ as the energy density of a higgs-type field $\varphi(x)$ (with an extra factor $\frac{1}{f_\pi}$ as a normalization factor),
which should include both the energy density of quark fields $\varrho_{q}$ and the energy density of gluon fields $\varrho_{g}$
(with the factor $\frac{1}{(m_u + m_d)}$ as a normalization factor);
and, in (\ref{fpion-on-V})
we treat the pion transition matrix element ${\mathcal M}_{\pi\rightarrow |0\rangle} \equiv \langle0|\bar{q}\gamma^0\gamma^5 q |\pi\rangle $ being proportional to $\Phi_{\bar{q}q}(x=0) $ ( i.e., the value of wavefunction $\Phi_{\bar{q}q}(x)$ at $x=0$ point
in the coordinate space), thus being inverse to the characteristic size $L$ (or the volume $V=L^3$) of pion by imposing a Gaussian type wavefunction in a constituent quark model scheme.
That will mean, the relation (\ref{mpion-on-mq}) in ChPT can be embedding in our model (\ref{mpion-on-rhoV}) by introducing assumptions in  (\ref{VEV-on-rho},\ref{fpion-on-V}).

\subsection{Why compact $N$-quark Hadrons Disappear?}\label{disappear-due-to-heavy}

By inserting $m_{u,d}$, $b^{J,L}_{N}$ and $a^{J,L}_{N}$ defined in (\ref{mu-and-md},\ref{bJLN-coefficient},\ref{aJLN-coefficient}) into Eq. (\ref{mass-rule}),
we can get the masses of a compact $4$-quark hadron with quantum numbers $J=0$, $L=0$
\ba
M^{J=0,L=0}_{N=4}(qq\bar{q}\bar{q})  &\simeq&     4^2 \cdot  4 \cdot (4-1) \cdot 15 \,MeV =  2880  \,MeV, \\
M^{J=0,L=0}_{N=4}(\{qq\bar{q}\bar{c},qq\bar{c}\bar{c},qc\bar{c}\bar{c},cc\bar{c}\bar{c}\})
&\simeq&
2880  \,MeV+ 1270\cdot\{1,2,3,4\} \,MeV \\
&=& \{ 4150,5420,6690,7960 \}   \,MeV, \\
M^{J=0,L=0}_{N=4}(\{qq\bar{q}\bar{b},qq\bar{b}\bar{b},qb\bar{b}\bar{b},bb\bar{b}\bar{b}\})
&\simeq&
2880  \,MeV+ 4200\cdot\{1,2,3,4\} \,MeV \\
&=&  \{ 7080,11280,15480,19680  \}    \,MeV,
\ea
and
the masses of a compact $5$-quark hadron with quantum numbers $J=\frac{1}{2}$, $L=0$  as
\ba
M^{J=\frac{1}{2},L=0}_{N=5}(qqqq\bar{q})   &\simeq&      5^2 \cdot  5 \cdot (5-1) \cdot 15 \,MeV   = 7500  \,MeV,\\
M^{J=\frac{1}{2},L=0}_{N=5}(\{qqqq\bar{c},qqqc\bar{c},qqcc\bar{c},qccc\bar{c},cccc\bar{c}\})
&\simeq&
7500  \,MeV +1270 \cdot\{1,2,3,4,5\}  \,MeV  \\
&=&  \{8770,10040,11310,12580,13850\} \,MeV,\\
M^{J=\frac{1}{2},L=0}_{N=5}(\{qqqq\bar{b},qqqb\bar{b},qqbb\bar{b},qbbb\bar{b},bbbb\bar{b}\})
&\simeq&
7500  \,MeV +4200 \cdot\{1,2,3,4,5\}  \,MeV  \\
&=&  \{11700,15900,20100,24300,28500\} \,MeV;
\ea
and we can see the masses are rather larger than the results in constituent quark models \cite{report2007}.

By comparing with the electro-weak decays of the constituent quarks and the $q\bar{q}$ annihilation decays in the compact $N$-quark hadrons,
because the mass of one
compact $2N$-quark meson $|q^{N}\bar{q}^{N}\rangle$ (or, $3N$-quark baryon $|q^{3N}\rangle$)
will be much larger than $N$ compact $2$-quark mesons $|q\bar{q}\rangle$ (or, $N$ compact $3$-quark baryons $|qqq\rangle$),
the width of the strong decay processes $|q^{N}\bar{q}^{N}\rangle\rightarrow N|q\bar{q}\rangle$ (or, $|q^{3N}\rangle\rightarrow N|qqq\rangle$) will be very large.
This can be easily understood by noting that, both the $|q^{N}\bar{q}^{N}\rangle$ (or, $|q^{3N}\rangle$) and the $N|q\bar{q}\rangle$ (or, $N|qqq\rangle$)
are the eigenstates of the full Hamiltonian and the decay probability is proportional to the energy difference between the higher energy level and the lower energy level in the quantum mechanics perturbative theroy.
Therefore,
on one hand, it is difficult to discover compact $N$-quark hadrons due to the too large decay widths;
on the other hand, it is difficult to produce  compact $N$-quark hadrons on the colliders due to the smaller phase space of motion for the larger $N$ value;
these are just the reasons why there are rarely definite signals of compact $N$-quark hadrons so far.

Unlike the compact $2N$-quark mesons, all of which will be at last decay via the $q\bar{q}$ annihilations,
if there are other unknown mechanisms (e.g., in the high temperature, high density and high pressure environment) forbidding the decays of $|q^{3N}\rangle\rightarrow N|qqq\rangle$,
there will exist neutral stable compact $3N$-quark baryons (as all the charged ones have decayed to the neutral ones via the weak interaction) deposited in the core of stars
and the neutral stable ones will not be so easy to detected and discovered.

\subsection{Glueballs}\label{Glueballs}

With the so-called Cho-Duan-Ge decomposition \cite{glueball}, we can gauge independently decompose the 8 gluons in $SU(3)$ QCD
to 2 color-neutral binding gluons (also called ``neuron" or``neuton") and 6 (or three complex) colored valence gluons (also called  ``chromon" or ``coloron"),  then the 6 chromons can condense to glueballs (also called the ``chromoballs").
According to the mass spectrum hypothesis above, i.e., Eq. (\ref{mass-rule}),
\ba
M^{JL}_{N}  =    ( m_1 + m_2 +\ldots + m_N ) +  N^2 \cdot  N (N-1) \cdot a_{N}^{JL} + b_{N}^{JL},  \quad (\ref{mass-rule}) \non
\ea
with the same $b_{N}^{JL}$ and $a_{N}^{JL}$ values in (\ref{bJLN-coefficient}) and (\ref{aJLN-coefficient}), we can fit a nonzero ``current gluon" mass
in the condensate occurring case as
\ba
m_g  =   120 \,MeV , \label{current-gluon-mass}
\ea
by treating the unidentified particle $X(360)$ \cite{PDG-2020}\cite{X360}
(with the quantum numbers
$I^G (J^{PC} )$
not identified yet)
as a $2$-gluon glueball state $|g\bar{g}\rangle$
with the quantum numbers $I^G (J^{PC} ) = 0^+(0^{++})$,
by recalling that the quantum numbers of gluon are $I(J^P)= 0 (1^{-})$.
Moreover,
by setting
the number of color charge of gluons to be the same as quarks,
the $2$-gluon and $3$-gluon glueball states would have the masses
\ba
I^G (J^{PC} ) = 0^+(0^{++}),\quad  && M^{J=0,L=0}_{g\bar{g}}  \simeq  360 \,MeV,  \\
I^G (J^{PC} ) = 0^+(1^{+-}),\quad  && M^{J=1,L=0}_{g\bar{g}}  \simeq  970 \,MeV,  \\
I^G (J^{PC} ) = 0^+(2^{++}),\quad  && M^{J=2,L=0}_{g\bar{g}}  \simeq  2190\,MeV,
\ea
and
\ba
I (J^{P} ) = 0(1^{-}),\quad  && M^{J=1,L=0}_{ggg}  \simeq  1370 \,MeV,  \\
I (J^{P} ) = 0(2^{-}),\quad  && M^{J=2,L=0}_{ggg}  \simeq  1780 \,MeV,  \\
I (J^{P} ) = 0(3^{-}),\quad  && M^{J=3,L=0}_{ggg}  \simeq  2390 \,MeV.
\ea

Since the dynamically reversible transition processes $gg \leftrightarrow q \bar{q}$ are going on all the time,
it is most likely that
the $2$-gluon glueball states $|g\bar{g}\rangle$ will be mixed with the
quarkonium states $|q \bar{q}\rangle$
and they would not be distinctly discovered.
Nevertheless, the glueball states with total spin $J=2,3$ (and orbit angular momentum $L=0$) might be more pure, and they would be more expectable to search.
More glueball states with other $I^G (J^{PC} )$ quantum numbers (e.g., see Ref.~\cite{report2007}\cite{glueball}) are also allowed in our model.

\section{\label{Constituent-quarks}Constituent Quarks as Physical d.o.f}

What is an $N$-quark hadron?
For a hadron state with definite quantum numbers, we can write out the Fock expansion as
\ba
|hadron,P\rangle = h_2|P;p_1p_2\rangle+ h_3|P;p_1p_2p_3\rangle+....+...,
\ea
where $p_i$ is momentum of the $i$-th constituent, such as constituent quarks and the colored valence gluons but not the free current quarks and the free gluons;
and, the coefficients $h_i$ might be functions of many different parameters, such as the energy scale $\mu$.
Generally, each Fock state is allowed, and we do not know which one is the most dominant. For example, the parton picture can be seemed as the  $i=\infty$ Fock state; although the $i= 2$ Fock states are always seemed as the leading-order ones in a meson state, there are no absolutely sufficient reasons to ignore the $|\bar{q} q g\bar{g}\rangle$ states but only for the simplicity.
That means, a hadron state is ``defined to be" an $N$-quark state in the Fock expansion language.

Moreover, for each Fock state, besides of the quantum numbers (such as: the spins, the flavors, the colors, etc),
we do not know the dynamical details of the constituents, i.e., the masses and the interactions.
That means,
a constituent quark is indeed ``defined to be" a physical quasi-particle d.o.f
after we define the hadron as an $N$-quark system, rather than directly deduced from the first principle of QCD.

After we treat constituent quarks as physical quasi-particle d.o.f, i.e., the ``single physical quark" states $|1_{M}\rangle = \alpha^\dag  | \Omega\rangle$, as said in (\ref{physical-quark-is-mixed-state}) and the last two paragraphs in Sect. \ref{Physical-dof},
we can get the masses of constituent quarks.
If we just define the ordinary mesons as $2$-quark states and the ordinary baryons as $3$-quark states,
\ba
|meson,P\rangle =  |P;p_1p_2\rangle,\,
|baryon,P\rangle = |P;p_1p_2p_3\rangle ,
\ea
then the masses of constituent quarks would be about $\frac{1}{2} M_{meson}$ and $\frac{1}{3} M_{baryon}$, respectively.
Why are the constituent quark model so successful?
It is just because that
the total mass of the constituent quarks is the dominant part of the hadron mass, while the residual effective potential energy can be seemed as perturbation.
Following this approach, one can solve the wavefunction $ \langle 0 | \bar{q}(x)q(y)  |hadron\rangle$ after some suitable potential model for the interaction of constituent quarks is constructed.

\section{Conclusions}\label{conclusions}

In this paper we try to give a hypothesis on the mass spectrum of compact $N$-quark hadrons and glueballs in a classical field picture, which indicates that there would be a mass dependence on about $N^4$. We call our model ``bag-tube oscillation model", which can be seemed as a kind of combination of quark-bag model and flux-tube model. The large decay widths due to large masses might be the reason why the compact $N$-quark hadrons still disappear so far.

\section{Acknowledgements}

I am very grateful to
Dr. Jia-Jun WU at University of Chinese Academy of Sciences (UCAS) for useful comments,
and I am also very grateful to
Prof. Xin-Heng GUO at Beijing Normal University
and
Dr. Xing-Hua WU at Yulin Normal University
for guidance on field theories before.

%%%%%%%%%%%%%%%%%%%%%%%%%%%%%%%%%%%%%%%%%%%%%%%%%%%%%%%%%%%%%%%%%%%%%%%%%%%%%%%%%%%%%%%

\newpage

%%%%%%%%%%%%%%%%%%%%%%%%%%%%%%%%%%%%%%%%%%%%%%%%%%%%%%%%%%%%%%%%%%%%%%%%%%%%%%%%%%%%%%%
%%%%%%%%%%%%%%%%%%%%%%%%%%%%%%%%%%%%%%%%%%%%%%%%%%%%%%%%%%%%%%%%%%%%%%%%%%%%%%%%%%%%%%%

%\end{CJK*}
\newpage

\end{document}